\documentclass[twocolumn,amsmath,amssymb,pra]{revtex4}

\usepackage{graphicx}
\usepackage{dcolumn}
\usepackage{bm}

\begin{document}

\preprint{APS/123-QED}

\title{Comment on ``Trouble with the Lorentz Law of Force: Incompatibility with Special Relativity
and Momentum Conservation''}

\author{Pablo L. Saldanha}\email{saldanha@df.ufpe.br}
\affiliation{Departamento de F\'isica, Universidade Federal de Pernambuco, 50670-901, Recife, PE, Brazil}
\affiliation{Departamento de F\'isica, Universidade Federal de Minas
Gerais, Caixa Postal 702, 30161-970, Belo Horizonte, MG,
Brazil}



\maketitle

Mansuripur argued in \cite{mansuripur12} that the Lorentz force law
is incompatible with the special relativity theory and momentum
conservation, based on an example in which a magnetic dipole do not
suffer any torque from an electric charge in the system rest frame,
but in other frames such torque appears if the Lorentz law is used
to compute the forces on it. However, here we show that there is no
paradox if the ``hidden momentum'' of a magnetic dipole in the
presence of an external electric field is taken into account.

The electric charge from Mansuripur's example \cite{mansuripur12}
produces an electric field $\mathbf{E}=E\mathbf{\hat{z}}$ on the
(infinitely small) magnetic dipole  with dipole moment
$\mathbf{m}_0=m_0\mathbf{\hat{x}}$. Different models that use
current loops to represent a magnetic dipole predict that the dipole
acquires a ``hidden momentum''
$\mathbf{P}=\varepsilon_0\mathbf{m}_0\times\mathbf{E}$  in the
presence of the external electric field due to relativistic effects
on the moving charges of the loops \cite{penfield,vaidman90}. This
``hidden momentum'' is counterbalanced by the electromagnetic
momentum obtained from the integral of
$\varepsilon_0\mathbf{E}\times\mathbf{B}=\varepsilon_0\mathbf{E}\times(\mu_0\mathbf{H}+\mathbf{M})$
in the whole space. Since we have $\mathbf{H}=0$ and
$\mathbf{M}=\mathbf{m}_0\delta^3(\mathbf{r}-\mathbf{r}_0)$ in
Mansuripur's example \cite{mansuripur12}, $\mathbf{r}_0$ being the
dipole position, the system total momentum is zero.

Now let us consider the same system in a reference frame that moves
with velocity $\mathbf{V}=-V\mathbf{\hat{z}}$. The electric field in
the position of the magnetic dipole is the same as before and there
is no magnetic field. However, the magnetic dipole acquires an
electric dipole moment $\varepsilon_0m_0V\mathbf{\hat{y}}$, such
that there is a net torque
$\mathbf{T}=E\varepsilon_0m_0V\mathbf{\hat{x}}$ acting on it
\cite{mansuripur12}. But if we take the ``hidden momentum'' into
account, there is no inconsistency. In the new frame this momentum
is the same, $\mathbf{P}=-\varepsilon_0m_0E\mathbf{\hat{y}}$. If the
magnetic dipole is at the origin of the system of coordinates at the
origin of time, its position in the new frame is
$\mathbf{r}=Vt\mathbf{\hat{z}}$ and the angular momentum is
$\mathbf{L}=\mathbf{r}\times\mathbf{P}=\varepsilon_0m_0EVt\mathbf{\hat{x}}$,
such that $d\mathbf{L}/dt=\mathbf{T}$. The torque is equal to the
rate of change of the angular momentum caused by the movement of an
object with ``hidden momentum'', such that there is no angular
acceleration of the dipole and no paradox arises with the use of the
Lorentz law. There is a ``hidden angular momentum'' that increases
in time, but is counterbalanced by the electromagnetic angular
momentum such that the system total angular momentum is constant in
time.


The magnetic dipole moment of quantum systems like atoms and electrons, on the other hand, cannot be described by classical current loops. So
it is not possible to say if such objects have or not ``hidden
momentum'' in the presence of an electric field based on the classical arguments from Refs.
\cite{penfield,vaidman90}.

This issue is related to the Abraham-Minkowski debate about the
momentum of electromagnetic waves in material media
\cite{pfeifer07}. The eventual conclusion of the debate is that
there are many different ways for dividing the total energy-momentum
tensor of the system into electromagnetic and material parts,
corresponding to different expressions for the electromagnetic
momentum density, force, energy flux, etc., that lead to the same
experimental predictions \cite{penfield,pfeifer07}. In previous
works \cite{saldanha10,saldanha11} we have shown that if the
momentum and energy transferred to matter by an electromagnetic wave
are computed by the use of the Lorentz law of force, we must
consider the expression $\varepsilon_0\mathbf{E}\times\mathbf{B}$
for the electromagnetic momentum density,
$\mathbf{E}\times\mathbf{B}/\mu_0$ for the electromagnetic energy
flux and take the ``hidden momentum'' and a ``hidden energy flux''
into account to have energy and momentum conservation in different
situations, as well as an agreement with Balazs \textit{gedanken}
experiment \cite{balazs53}. Here we are confirming this fact. On the
other hand, with the use of the Einstein-Laub force one must
consider the expressions $\mathbf{E}\times\mathbf{H}/c^2$ for the
electromagnetic momentum density, $\mathbf{E}\times\mathbf{H}$ for
the electromagnetic energy flux and disregard the ``hidden
momentum''. Since both formulations give the same experimental
predictions when properly used, they are equally valid.

It is important to stress that the Lorentz force law can be written
in a covariant way, such that it is automatically compatible with
special relativity. Mansuripur's apparent paradox can be solved
based on this fact, as discussed in Refs. \cite{vanzella12,cross12,mcdonald12}. It is also worth
to mention that similar apparent paradoxes regarding the torque on
magnetic dipoles were discussed and solved many years ago \cite{vaidman90,bedford86,namias89}.


This work was supported by the Brazilian agencies CNPq and FACEPE.

\vspace{0.5cm}

\textit{Note added}: After the submission of the first version of this work
for publication, several solutions of Mansuripur's paradox were
posted on the Internet based on essentially  the same arguments that
we present here. See \cite{vanzella12,cross12,mcdonald12,griffiths12} to quote a few.

\end{document}